\documentclass[10pt,conference,compsocconf]{IEEEtran}
\IEEEoverridecommandlockouts
\usepackage{graphicx}
\usepackage{times,amsmath,epsfig}
\usepackage{mathrsfs}
\usepackage{amssymb}
\usepackage{times,amsmath,epsfig}
\usepackage{amsmath,graphicx,latexsym}
\usepackage{algorithm,algorithmic}
\usepackage{changebar}
\usepackage{verbatim}
\usepackage{subfigure,afterpage}
\usepackage{authblk}
\usepackage{soul}
\usepackage{graphicx}

\newtheorem{definition}{Definition}

\renewcommand{\comment} [1]{}

\begin{document}
%
\title{Privacy Preserving Driving Style Recognition}

\author{Nicholas Rizzo \thanks{This work is supported in part by the SUNY STEM Passport Program. Corresponding Author: Yuan Hong}, 
	Ethan Sprissler, Yuan Hong, and Sanjay Goel
	\\Department of Information Technology Management
	\\University at Albany, SUNY, Albany, NY 12222
	\\\{nrizzo, esprissler, hong, goel\}@albany.edu

}


%


\maketitle

\begin{abstract}

In order to better manage the premiums and encourage safe driving, many commercial insurance companies (e.g., Geico, Progressive) are providing options for their customers to install sensors on their vehicles which collect individual vehicle's traveling data. The driver's insurance is linked to his/her driving behavior. At the other end, through analyzing the historical traveling data from a large number of vehicles, the insurance company could build a classifier to predict a new driver's driving style: aggressive or defensive. However, collection of such vehicle traveling data explicitly breaches the drivers' personal privacy. To tackle such privacy concerns, this paper presents a privacy-preserving driving style recognition technique to securely predict aggressive and defensive drivers for the insurance company without compromising the privacy of all the participating parties. The insurance company cannot learn any private information from the vehicles, and vice-versa. Finally, the effectiveness and efficiency of the privacy-preserving driving style recognition technique are validated with experimental results.


\end{abstract}


%
\IEEEpeerreviewmaketitle

\section{Introduction}
\label{sec:intro}
One of the engineering innovations that will have a transformative impact on the society over the next few decades is the Connected Vehicles initiative \cite{cvehicle}. The connectivity is enabled through the use of wireless communication over a dedicated spectrum to create local ad hoc networks of vehicles that are then able to communicate with other vehicles in its neighborhood, as well as with traffic infrastructure. The communication network provides opportunity for mobile devices or inertial sensors mounted in the vehicles to collect data regarding vehicle travel, driver behavior, and location data \cite{Ly13,Zhang13}. This data can be processed for implementing automatic collision avoidance, optimizing traffic in real time, and planning for infrastructure needs. While, these data sets can improve safety, reduce emissions, and reduce driver wait time, they can also be used to uncover sensitive personal information. For instance, location data can be used in criminal investigations, driver behavior can be used to determine fault in accidents, and a person's lifestyle choices can be revealed by processing the data with mapping software. The privacy challenge needs to be addressed with a careful balance between the utility of the collected data and the protection of personal information as well as corporate proprietary information.

The insurance industry can leverage vehicle travel data to determine their premiums by correlating driver behavior and accidents. For instance, vehicle data can be used to create a classification system that can classify drivers as defensive or aggressive \cite{Ly13}. In order to better manage their premiums and also encourage safe driving, many commercial insurance companies (e.g., GEICO and Progressive) are providing options for customers to install sensors on their vehicles that can collect the vehicle's operational parameters (such as braking, acceleration, speed, etc.) \cite{insure}. Their insurance is linked to their driving behavior. At the other end, through analyzing the historical travel data from a large number of vehicles, the insurance company could build a classifier to determine the thresholds for aggressive vs. defensive driving. This would allow the companies to better understand the risks associated with each of their drivers and be able to tailor premiums based on the risks. Insurance companies could also give drivers ``report cards'' to help drivers better understand their own driving habits. However, collection of the vehicle traveling data to make this happen explicitly breaches the drivers' personal privacy. For instance, the accelerating \& braking data, trips, turning behaviors, risky driving hours, and even the geographical locations the driver can be inferred from the data violating their personal privacy. The goal of this research is to securely build a classifier that accurately predicts any given driver's driving style (aggressive or defensive) without compromising any participating party's privacy.

There are two phases in driving style recognition: (1) building a classifier with vehicles' historical traveling data; (2) predicting the driving style for new drivers using the classifier. 

\begin{enumerate}
	\item In phase (1), some drivers' historical vehicle traveling data 
	are analyzed to train the classifier. As part of this process, each record has a class label applied to it, the characteristics of an aggressively labeled driver would be based on the records of individuals who had received speeding tickets or been involved in an accident, while a defensive driver would be an individual who had not been in such situations. 
	Each driver privately holds a record of his/her vehicle traveling data with the attributes such as average acceleration ($m/s^2$), average deceleration in the braking events ($m/s^2$), average turning (degrees), \# of risky driving hours, and \# of trips \cite{Ly13}. In reality, both of the driver and insurance company know the driver's class label in the training data from the traffic tickets, reported accidents, etc. Therefore, such multiparty classifier training process has a hybrid scenario of data partitions: first, all the drivers' records are horizontally partitioned -- each driver privately holds a record, including the class label \cite{JSVaidyaVLDBJ08}; second, the overall data (attributes) is also vertically partitioned \cite{JSVaidyaVLDBJ08} into two shares -- the insurance company holds a share (the attribute class label) while all the drivers jointly holds both shares, including the class label. Finally, the output of phase (1) is a trained classifier, privately held by the insurance company.
	
	\item In phase (2), the insurance company privately holds its classifier whereas the new driver privately holds his/her vehicle traveling data. Since the two parties privately own different attributes, it is a vertical partition case in each individual driving style recognition. Then, they jointly predict the class label (aggressive or defensive) with their private inputs.
	
\end{enumerate}

 Since the data partitions in the previous two phases are mixed with both horizontal and vertical partitions, the prior works on privacy-preserving classification \cite{Lindell00,JSVaidyaVLDBJ08,JSVaidyaKAIS08,XiaoHLS05} are not directly applicable to this research problem.

\subsection{Research Contributions}

In this paper, we develop privacy-preserving techniques for two phases of driving style recognition based on decision tree induction \cite{ID3}. More specifically, in phase (1), each driver only knows its vehicle traveling data (a record) and its class label; the insurance company only knows all the drivers' class labels, and the final output -- a decision tree. In phase (2), each new driver only knows its vehicle traveling data (a record) and learns nothing but the driving style recognition result; the insurance company only knows its input (the decision tree) and also learns nothing but the driving style recognition result. Thus, the main contributions are summarized as below:

\begin{enumerate} 
	
	\item We propose two secure communication protocols under secure multiparty computation (SMC) \cite{Yao86, Goldreich} to implement the two phases of driving style recognition (based on decision trees) for the participating parties without private information disclosure.
	\item We analyze the privacy risks for all the participating parties in the secure communication protocols for both classifier training and driving style prediction. 
	\item We experimentally validate the performance of our proposed approaches on the synthetic datasets generated following the format of data collected in \cite{Ly13}.
\end{enumerate}

The remainder of this paper is organized as follows. We first briefly review the related work in Section \ref{sec:related}. Then, we present the algorithms and demonstrate the experimental results in Sections \ref{sec:protocol} and \ref{sec:exp} respectively. Finally, we conclude this paper and discuss the future work in Section
\ref{sec:concl}.


\section{Related Work}
\label{sec:related}

Vehicle traveling data has been commonly collected for analysis in Intelligent Transportation Systems (ITS). Ly et al. \cite{Ly13} demonstrated a methodology to collect vehicle data (e.g., accelerating, braking, turning) using inertial sensors. Hull et al. \cite{Hull06} collected the vehicle data with his CarTel system. There has also been expansive work into adaptive cruise control, which uses prediction algorithms to adapt to the road based on curve patterns \cite{Zhang13}. Moreover, Rass et al. \cite{RassFK08} formulated a system to provide feedback on the driving habits. Also, other research has focused on driver modeling \& evaluation \cite{MiyajimaUNAKT11, Kishimoto08} and maneuver recognition \cite{Sathyanarayana12}. While there has been a variety of ventures into many ITS applications, only a surprising few of these tackled such arising privacy concerns, including: Hoh et al. \cite{Hoh08} proposed using the virtual trip lines to monitor traffic conditions while preserving privacy. Checkoway et al. \cite{CheckowayMKASSKCRK11} examined the attack vectors for hackers to infiltrate vehicle through its Electronic Control Unit. Han et al. evaluated authentication methods for securely integrating mobile devices in vehicular networks \cite{HanPS13}. To the best of our knowledge, privacy risks in driving style recognition have not been systematically studied. 


Privacy-preserving schemes are generally developed based on data transformation and/or secure computation. The former one transforms the original data to a privacy-compliant format and minimizes the utility loss in the process of data transformation (e.g., k-anonymity \cite{Sweeney02}, differential privacy \cite{DworkMNSTcc06}). The latter makes two or more parties jointly compute a function possible without revealing private data to each other (formally defined as Secure Multiparty Computation \cite{Yao86}). The function can be as simple as sum or as complex as big data analysis/mining \cite{Agrawal00,Lindell00,HongThesis}. Several researchers have addressed the privacy concerns in other contexts, such as classification \cite{Lindell00,XiaoHLS05,JSVaidyaVLDBJ08}, location based services \cite{GedikL05}, 
search engine queries \cite{HongCikm09,HongEDBT12,HongTDSC15}, scheduling \cite{HongTabu}, transportation \cite{HongJCS12}, and smart grid \cite{HongIJER15}. Following a similar line of research, we develop a privacy-preserving driving style recognition technique that can analyze the vehicle traveling data (e.g., an insurance company) without breaching participating parties' privacy.

\section{Secure Communication Protocols}
\label{sec:protocol}
In this paper, we assume that the adversaries are semi-honest. The semi-honest model in Secure Multiparty Computation (SMC) \cite{Yao86, Goldreich} defines that the adversaries are honest to follow the a given protocol, but are curious to infer private information from each other. Two secure communication protocols will be given for two phases of driving style recognition in Sections \ref{sec:training} and \ref{sec:securesum} respectively. 

\subsection{Phase (1): Privacy Preserving Classifier Training}
\label{sec:training}

In phase (1) of driving style recognition, a set of drivers and the insurance company jointly derive a decision tree based on the drivers' vehicle traveling data and class labels (aggressive or defensive). In this scenario, we have:

\begin{itemize}
	\item Every driver's class label in the training data is known by both the driver and the insurance company. Since the ``Aggressive'' class label in the historical data can be derived from  the ``traffic tickets or accidents'' which is indeed known by both parties in real world.
	
	\item All drivers and the insurance company know the name of every attribute in the dataset, such as Acceleration Events (\#), Average Acceleration ($m/s^2$), Braking Events (\#), and Average Braking ($m/s^2$). Insurance company initializes the attributes in the sensors which are also known to the drivers in real world.
	
	\item Every attribute has a threshold to divide the values into two categories: ``less than the threshold'' or ``no less than the threshold''. For example, the average deceleration when braking can be less than a threshold $4.9 m/s^2$ or no less than a threshold. Such split is used to determine the branches of a node on the decision tree. We assume that every attribute's threshold is known to all the drivers and the insurance company (e.g., thresholds of speed/mileage/acceleration/braking can be available as public information).
	
\end{itemize}



Note that our privacy-preserving decision tree training is extended from the ID3 Algorithm \cite{ID3,Lindell00}, which iteratively finds the best attribute to split values based on its threshold (as the current node of the tree) by comparing the entropy or information gain of all the remaining attributes in the classification results. In this paper, we choose the entropy as the measure of uncertainty in the threshold based split $H=-\sum_{x\in X}p(x)log p(x)$ where $X$ is the set of classes (``Aggressive'' or ``Defensive'') and $p(x)$ is the proportion of the number of elements in class $x$ to the number of elements in all the data. Therefore, in a distributed manner, each driver/vehicle owns a record and they should securely sum their shares for every $p(x)$. In this section, we first present the Secure Sum algorithm which is iteratively invoked by the protocol of privacy-preserving classifier training. 

\subsubsection{Secure Sum}
\label{sec:securesum}



The secure sum algorithm is developed using Homomorphic Cryptosystem (e.g., Paillier \cite{Paillier99}). It begins by having the insurance company $I$ generate a key pair: a public key $pk$ and a private key $sk$. The insurance company (party $I$) then sends the public key $pk$ to all $m$ drivers, denoted as $D_1,\dots, D_m$. $D_1$ then encrypts its share and passes along the encrypted data to the next driver (w.l.o.g, say $D_2$). $D_2$ then computes their encrypted sum to the previous number and this is passed through all the $m$ drivers in the circuit. After this, the encrypted sum is passed back to the insurance company who decrypts it with the private key $sk$ to get the sum. As shown in Algorithm \ref{algo:secsum}, all the parties' data remains private while only allowing the insurance company to obtain the sum.

\begin{algorithm}[!h]
	\begin{algorithmic}[1]
		
		\renewcommand{\algorithmicrequire}{\textbf{Input:}}
		\renewcommand{\algorithmicensure}{\textbf{Output:}}
		
		\REQUIRE $m$ drivers' share of $p(x)$: $p(x)_1,\dots, p(x)_m$ 
		\ENSURE insurance company $I$ learns $p(x)=\sum_{j=1}^mp(x)_j$
		
		\STATE $I$ generates a pair of public-private key $(pk, sk)$ and sends the public key $pk$ to $P_1,\dots,P_n$
		
		\FOR{$i=1,\dots, m$}
		%
		%
		
		\STATE $D_i$ encrypts $p(x)_i$ using $pk$ to get $E[p(x)_i]$, and computes $E[\sum_{j=1}^ip(x)_j]=\prod_{j=1}^{i-1}E[p(x)_j]*E[p(x)_i]$ (ensured by Homomorphic Property \cite{Paillier99})

		\STATE $D_i$ sends $E[\sum_{j=1}^ip(x)_j]$ to the next party $P_{i+1}$ (if $i=m$, the next party is $I$)
		
		\ENDFOR
		
		\STATE $I$ decrypts $E[\sum_{j=1}^mp(x)_j]$ with its private key $sk$ to obtain $\sum_{j=1}^mp(x)_j$
		
	\end{algorithmic}
	\caption{Secure Sum}\label{algo:secsum}
\end{algorithm}

\subsubsection{Secure Communication Protocol}\label{sec:pptrain}


In the secure communication protocol for classifier training, the insurance company $I$ repeatedly finds the best attribute (that has the smallest entropy $H$; viz. lowest uncertainty) as the current node during the construction of the tree. 
Note that the key value $p(x)$ in the entropy $H=-\sum_{\forall x\in X}p(x)\log p(x)$ is split into $m$ shares $p(x)_1,\dots, p(x)_m$, held by $m$ vehicles respectively. In each iteration, each of the remaining attributes' $p(x)$ is securely summed by all $m$ vehicles and the insurance company $I$ (Algorithm \ref{algo:secsum}). This use of the secure sum ensures that $I$ is only ever able to learn the $p(x)$ of each attribute while preserving the privacy of each of the $m$ vehicles. 

In turn, the only information that is learned by each of $m$ vehicles is the $pk$ that is sent by the insurance company $I$. The purpose of this approach is to provide a mechanism to securely sum the data needed to compute the entropy values for each attribute while obfuscating all vehicle identifying characteristics in $p(x)_1,\dots,p(x)_m$. In addition, the use of the secure sum ensures that no other parties will be able to uncover the identity of any vehicle or the vehicle's collected data.

In the final phase of this algorithm, $I$ locally computes the entropy $H$ of all the remaining attributes, and selects the attribute with the smallest entropy value as the current node. This process is performed iteratively until all the leaf nodes of the decision tree have $p(x)=1$. The details of the secure communication protocol is given in Algorithm \ref{algo:pptrain}

\begin{algorithm}[!h]
    \begin{algorithmic}[1]

        \renewcommand{\algorithmicrequire}{\textbf{Input:}}
        \renewcommand{\algorithmicensure}{\textbf{Output:}}

        \REQUIRE $m$ is the total number of drivers/vehicles
        \ENSURE Decision tree $T$

        \WHILE{existing a leaf node's $p(x)!=1$}

        \FOR{$i = 1, \dots, m$}

        \STATE Driver $D_i$ computes the share of $p(x)$ for all the remaining attributes 

        \ENDFOR

        \STATE Securely sum shares of $p(x)$ (Algorithm \ref{algo:secsum}) for all the remaining attributes (only party $I$ knows that)

        \STATE $I$ computes entropy $H=-\sum_{i=1}p(x)log p(x)$ for all the remaining attributes


        \STATE $I$ selects the best attribute (smallest entropy) as the root or leaf node (leaf node is added along the branch of the tree with $p(x)!=1$)


        \ENDWHILE

    \end{algorithmic}
    \caption{Privacy Preserving Classifier Training}\label{algo:pptrain}
\end{algorithm}


The final product of this algorithm consists of a decision tree which contains a set of the attributes from the dataset. This tree outlines the pathways which represent determinable outcomes based on any given driver's collection of vehicle traveling data. This tree will be used 
to help securely predict a given new driver as either aggressive or defensive.

\comment{
\textbf{Please discuss the following in details: while securely training the decision tree, insurance company $I$ only learns every attribute's $p(x)$ in each iteration. All $n$ vehicle only learns the public key $pk$ generated by the insurance company $I$. No private information is revealed to other parties.}
}

\subsection{Phase (2): Privacy-preserving Driving Style Recognition}
\label{sec:dsr}



The insurance company has the means to classify driving style/behavior for given drivers, but has an equal interest in keeping its decision tree $T$ private. 
To this end, we develop another secure communication protocol to predict the driving styles without sharing information between the insurance company and the new driver, detailed below. 


\begin{definition} [Aggressive Path in Decision Tree $T$]
	is defined as a path in the decision tree $T$ that leads to the class of aggressive driving. 
\end{definition}

\begin{figure}[!htb]
	\centering \subfigure[Decision Tree]{
		\includegraphics[angle=0, width=1\linewidth]{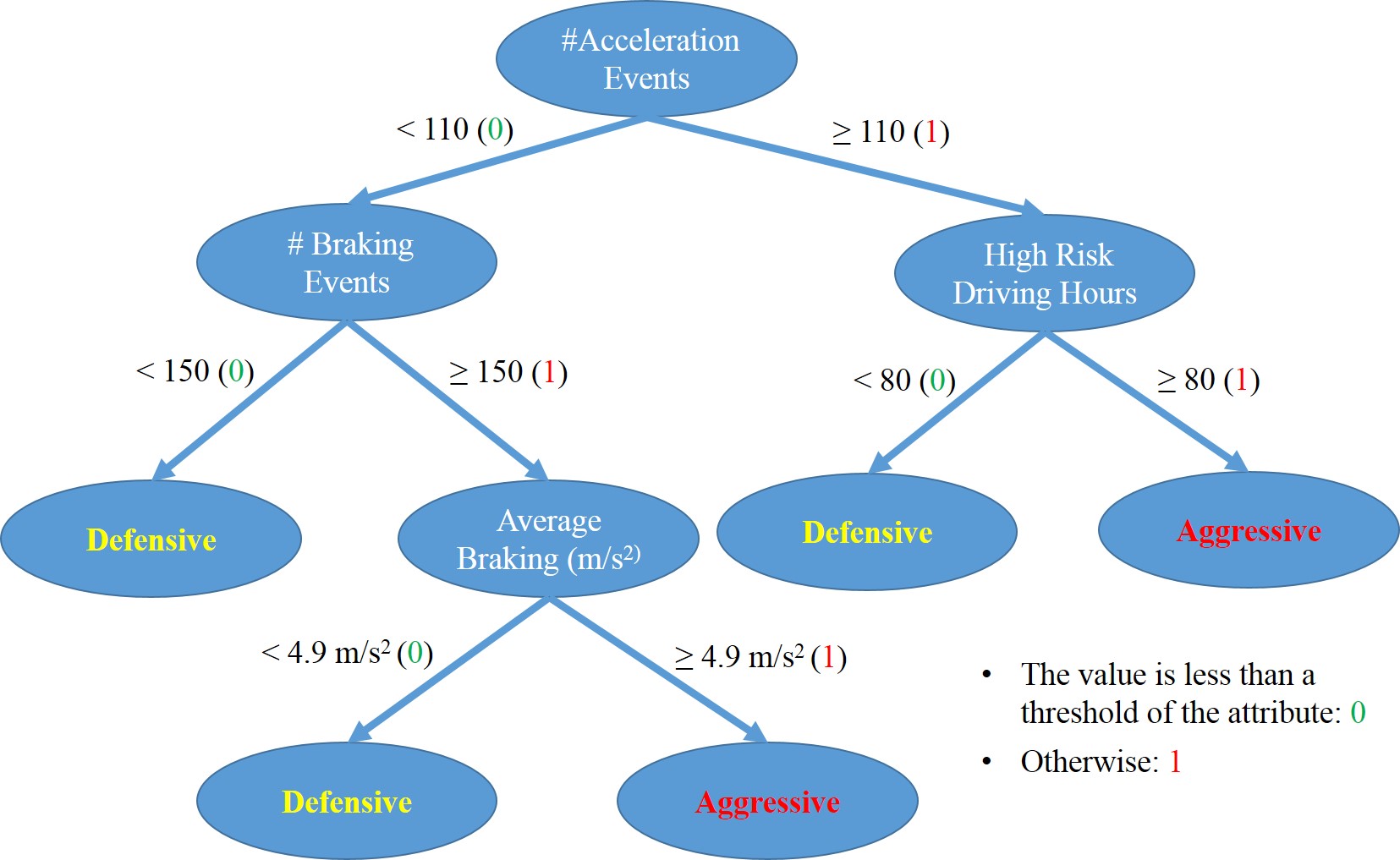}
		\label{fig:tree} } \subfigure[Aggressive and Defensive Paths]{
		\includegraphics[angle=0, width=1\linewidth]{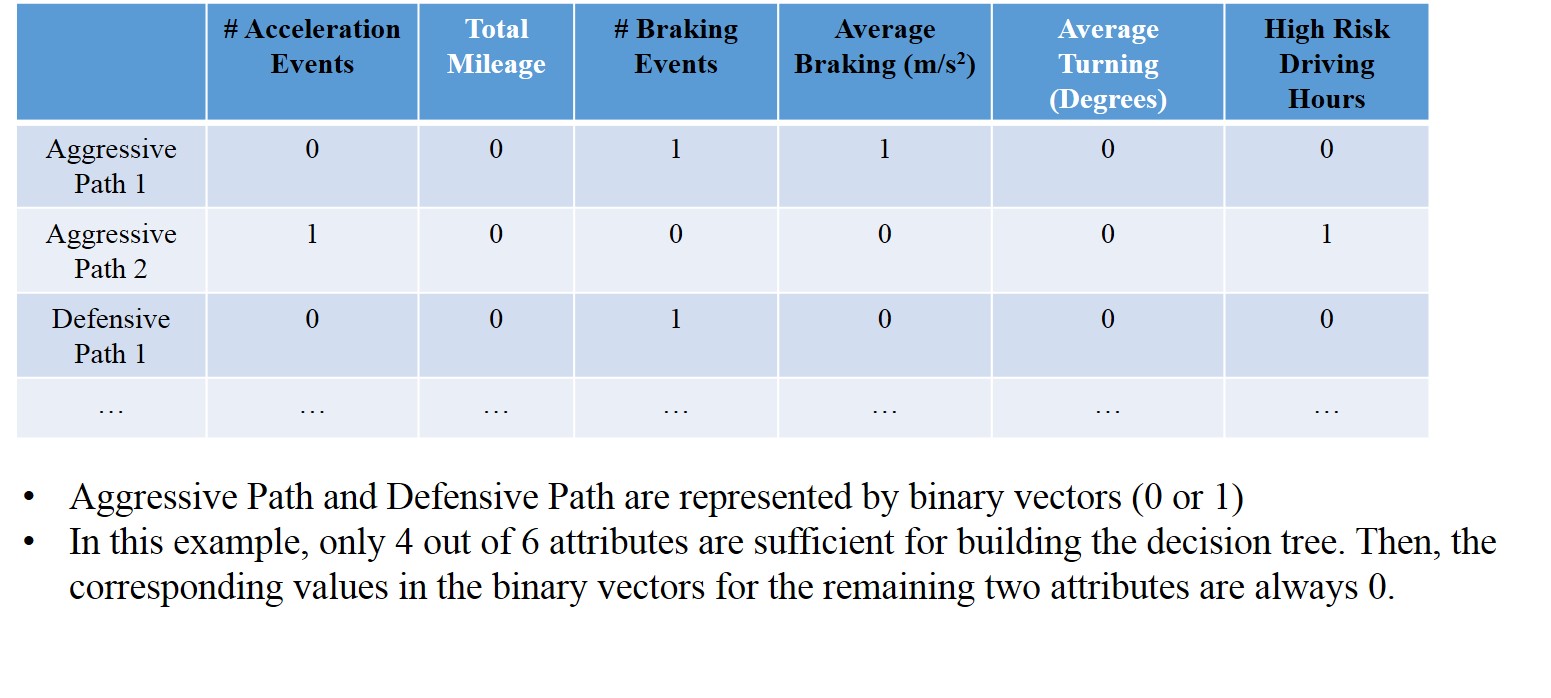}
		\label{fig:path} } \caption[Optional caption for list
	of figures]{An Example of Decision Tree and Aggressive/Defensive Paths} \label{fig:exp}
\end{figure}

Letting $|T|$ be the number of aggressive paths in $T$, with all the thresholds of the attributes in $T$, each aggressive path can be represented by an $n$-digit binary vector:

\begin{equation}
\forall i\in [1,|T|], \vec{c_{i}}=[c_{i1},c_{i2},\dots, c_{in}]
\end{equation}

where $\forall c_{ij}\in \{0,1\}$. Note that, in the decision tree $T$, $c_{ij}=1$ means the child value of the $j^{th}$ attribute (out of all $n$ attributes in total) along the aggressive path $\vec{c_i}$ exceeds the threshold value; otherwise, $c_{ij}=0$. For instance, in Figure \ref{fig:exp}, there are six attributes in total used for training decision tree, and four of them are utilized to build the tree $T$ (as shown in Figure \ref{fig:tree}). Two paths in $T$: ``\# of acceleration events $< 110$ $\longrightarrow$ \# of braking events $\geq 150$ $\longrightarrow$ average braking ($m/s^2$) $\geq 4.9 m/s^2$'' and ``\# of acceleration Events $\geq 110$ $\longrightarrow$ high risk driving hours $\geq 80$'' can lead to ``Aggressive''. Then, they can be represented as two binary vectors $(0,0,1,1,0,0)$ and $(1,0,0,0,0,1)$ respectively. Other paths in $T$ are simply considered as ``Defensive Paths'' which can be also represented as $n$-digit binary vector in a similar way.

The insurance company $I$ now owns the decision tree $T$, the number of aggressive paths $|T|$, as well as the path(s) that identify aggressive driving behavior $\vec{c_{|T|}}$. Since the insurance company privately possesses such information, the aggressive paths must be encrypted for computation. Specifically, $I$ generates a public/private key pair $(pk,sk)$ based on the Paillier cryptosystem \cite{Paillier99}. $I$ encrypts the aggressive paths from the decision tree and the total number of aggressive paths $|T|$ as well as the inner products of all the aggressive paths (which is the total number of ``1'' in each binary vector) using the public key $pk$ such that: $E(\vec{c_1})$, \dots, $E(\vec{c_{|T|}})$ and $E(\vec{c_1} \cdot \vec{c_1})$, \dots, $E(\vec{c_{|T|}} \cdot \vec{c_{|T|}})$ are both then transmitted along with $pk$ to the new driver/vehicle $D$. 

At the other end, similar to the aggressive/defensive paths, the new driver $D$ privately holds a:


\begin{definition} [Vehicle Traveling Vector $\vec{v}$]
is an $n$-digit binary vector: $\vec{v}=[v_1,v_2,\dots, v_n]$ with $0$ representing the attributes with values below the threshold, and otherwise $1$. 
\end{definition}


After receiving $pk$ and the ciphertexts from the insurance company $I$, the driver/vehicle $D$ then securely computes the following scalar products with the ciphertexts and its vector $\vec{v}$: 



\begin{equation}
\forall i\in [1,|T|], E(\vec{c_i}\cdot \vec{v}) =
E(c_{i1})^{v_{1}}*E(c_{i2})^{v_{2}}*\dots*E(c_{in})^{v_{n}}
\label{eq:scalar}
\end{equation}


Then, driver/vehicle $D$ encrypts $\vec{v}$ and computes:

\begin{equation}
\forall i\in[1,|T|], E(\vec{c_i}\cdot \vec{v}-\vec{c_i}\cdot \vec{c_i}) = \frac{E(\vec{c_i}\cdot \vec{v})}{E(\vec{c_i}\cdot \vec{c_i})}
\label{eq:onepath}
\end{equation}



If any of $\forall i\in[1,|T|], \vec{c_i}\cdot \vec{v}-\vec{c_i}\cdot \vec{c_i}$ equals $0$, the vehicle traveling vector $\vec{v}$ would match the corresponding aggressive path, and the driver $D$ is predicted as an aggressive driver. If all of $\forall i\in[1,|T|], \vec{c_i}\cdot \vec{v}-\vec{c_i}\cdot \vec{c_i}$ are non zero, the vehicle traveling vector $\vec{v}$ would not match any aggressive path, and the driver $D$ can be predicted as a defensive driver. To minimize information disclosure, the driver permutes all the ciphertexts $\forall i\in[1,|T|], E(\vec{c_i}\cdot \vec{v}-\vec{c_i}\cdot \vec{c_i})$ and send them to the insurance company $I$ one by one, and $I$ decrypts a ciphertext immediately. As long as a $0$ is found, conclude $D$ as an aggressive driver and terminate the protocol (no more ciphertext will be sent). If no $0$ is found after examining all $|T|$ results, then conclude $D$ as an defensive driver. $I$ can share the classification result to $D$ if necessary.


\comment{

\textbf{Also, needs more details for describing why one of $(c_1\cdot v-c_1\cdot c_1)$, \dots, $(c_{|T|}\cdot v-c_{|T|}\cdot c_{|T|})=0$ leads to aggressive
...and all non zero leads to defensive. Describes why number of ``1'' in the aggressive path equals the scalar product $c_1\cdot v$ make $v$ aggressive.You can
use an example to illustrate it}

}


\begin{algorithm}[!h]
    \begin{algorithmic}[1]

        \renewcommand{\algorithmicrequire}{\textbf{Input:}}
        \renewcommand{\algorithmicensure}{\textbf{Output:}}

        \REQUIRE Insurance company $I$ and a new driver $D$;
        \\~~~ $D$'s vehicle traveling vector $\vec{v}$;
        \\~~~ $T$ represents the complete decision tree;
        \\~~~ The number of aggressive paths $|T|$;
        \\~~~ All the aggressive paths $\vec{c_1},\dots, \vec{c_{|T|}}$
        \ENSURE The new driver is aggressive or not
        
        \COMMENT{A random nonce is generated for every single encryption}

        \STATE Party $I$ generates a public/private key pair based on Paillier Cryptosystem $(pk, sk)$

        \STATE $I$ encrypts $\vec{c_1},\dots, \vec{c_{|T|}}$, $|T|$, and inner products $\vec{c_1}\cdot \vec{c_1}$, \dots, $\vec{c_{|T|}}\cdot \vec{c_{|T|}}$ using $pk$ to obtain
	 $E(\vec{c_1}),\dots,E(\vec{c_{|T|}})$, $E(\vec{c_1}\cdot \vec{c_1})$, \dots, $E(\vec{c_{|T|}}\cdot \vec{c_{|T|}})$, and sends the ciphertexts and $pk$ to the driver $D$

        \STATE $D$ encrypts $\vec{v}$ and computes the ciphertexts of $|T|$ scalar products $E(\vec{c_1}\cdot \vec{v})$, \dots, $E(\vec{c_{|T|}}\cdot \vec{v})$ using Equation \ref{eq:scalar}


        \STATE $V$ computes $E(\vec{c_1}\cdot \vec{v}- \vec{c_1}\cdot \vec{c_1})$, \dots, $E(\vec{c_{|T|}}\cdot \vec{v}-\vec{c_{|T|}}\cdot \vec{c_{|T|}})$ using Equation \ref{eq:onepath} and permutes them
        
        

		\FOR{$i=1,\dots, |T|$}
		
		\STATE $D$ sends the permuted ciphertext $E(\vec{c_i}\cdot \vec{v}- \vec{c_i}\cdot \vec{c_i})$ to $I$
		
		\STATE $I$ decrypts the current $E(\vec{c_i}\cdot \vec{v}- \vec{c_i}\cdot \vec{c_i})$ using its private key $sk$ to get $\vec{c_i}\cdot \vec{v}- \vec{c_i}\cdot \vec{c_i}$
		
		\IF{$\vec{c_i}\cdot \vec{v}- \vec{c_i}\cdot \vec{c_i}=0$}
		
		\STATE $D$ is an aggressive driver and terminate the algorithm
		
		\ENDIF
		
		\ENDFOR

		\IF{$\forall i\in[1,|T|], \vec{c_i}\cdot \vec{v}- \vec{c_i}\cdot \vec{c_i}\ne 0$}

		\STATE $D$ is a defensive driver

		\ENDIF

    \end{algorithmic}
    \caption{Privacy Preserving Driving Style Recognition}\label{algo:ppreco}
\end{algorithm}

Upon completion, each driver has the ability to access his or her computed rating of either aggressive or defensive. $I$ has sole possession of the decision tree $T$ developed from training data and is the only party which is able to view all of the pathways which lead to an aggressive classification. The driver $D$, on the other hand, is the only party able to access $\vec{v}$, keeping specific driving behavioral data private from $I$. Ultimately, both parties will have access to the computed classification result. However, the insurance company can only infer some trivial information from $D$ such as how many values in $\vec{v}$ has met or exceeded the corresponding attributes' threshold. 


\section{Experimental Results}
\label{sec:exp}
We implemented application of privacy-preserving driving style recognition in Java on a PC with AMD FX-4350 4.55 GHZ CPU and 16G RAM. 
Synthetic datasets are generated falling into a similar value range as \cite{Ly13}. 
7500 drivers' vehicle traveling data are generated for training classifier (with class labels in the training data) while 2500 drivers' traveling data are generated for predicting the driving style by the classifier (without class labels in the dataset). These records 
simulated driving activity over a 6-month period and featured 9 attributes: 

\begin{itemize}
\setlength\itemsep{0.1em}
	\item total number of trips taken 
	\item total mileage driven 
	\item the number of acceleration events 
	\item the average amount of acceleration 
	\item the number of braking events 
	\item the average deceleration when braking 
	\item the average number of degrees turned 
	\item the total number of hard braking events 
	\item the hours driven in the highest risk time (0 to 4 AM)
		
\end{itemize}


The cryptographic keys are generated with lengths of 512 and 1024-bit using Paillier Homomorphic Cryptosystem \cite{Paillier99}.

For examining the computational costs, we tested the overall runtime of the protocols including encryption, computation and decryption. For examining the communication overheads, we tested the overall bandwidth consumption, which is equivalent to the overall size of the ciphertexts and plaintexts to be transmitted among all the distributed parties in the protocols. Due to the novelty of the data partition scenario and protocols devised for driving style prediction, there are no available benchmarks to compare against.




 \subsection{Classifier Training}

\begin{figure}[!h]
	\centering \subfigure[Computational Cost]{
		\includegraphics[angle=0, width=0.8\linewidth]{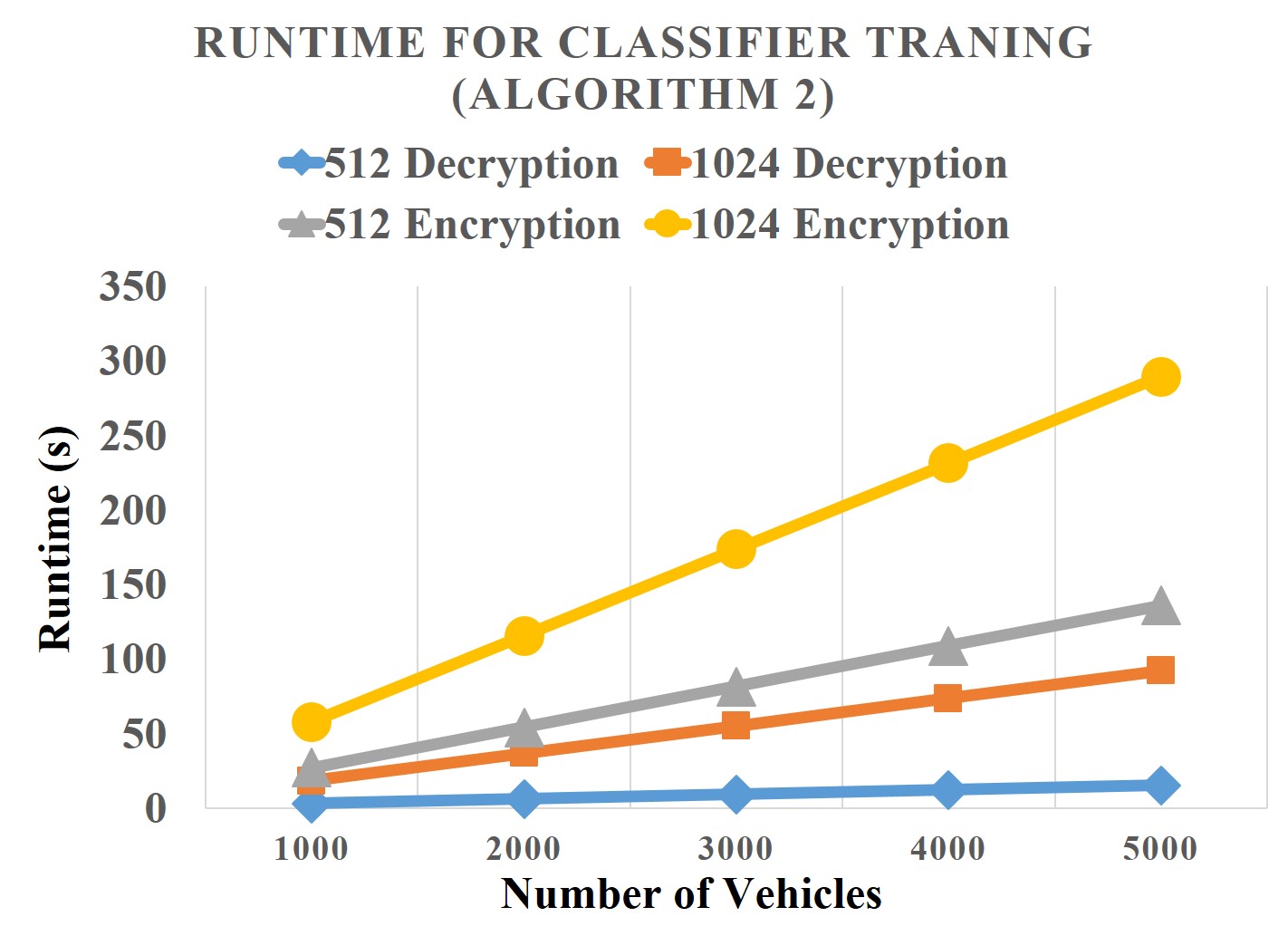}
		\label{fig:algm2time} } \subfigure[Communication Overheads]{
		\includegraphics[angle=0, width=0.8\linewidth]{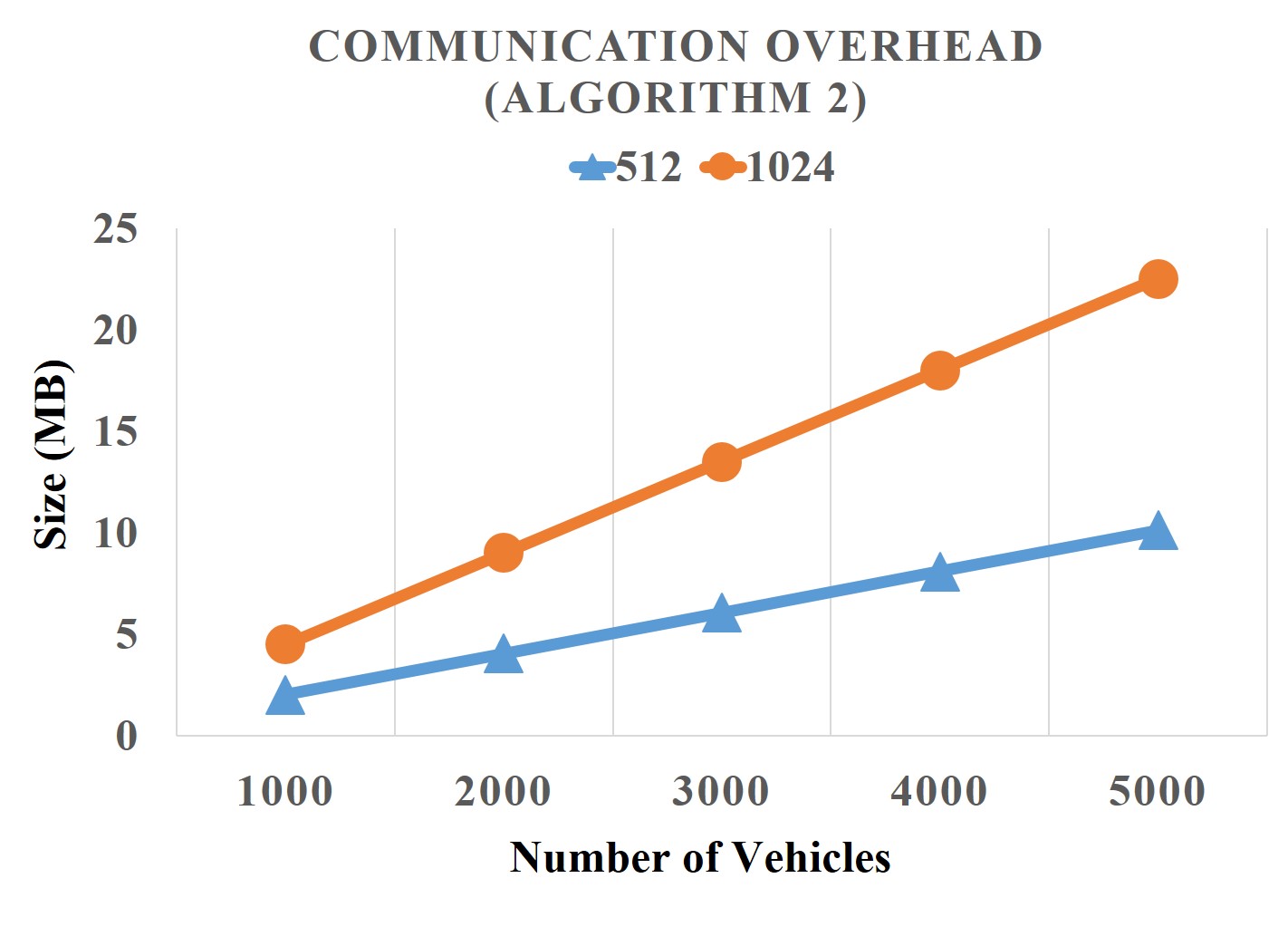}
		\label{fig:algm2size} } \caption[Optional caption for list
	of figures]{Privacy Preserving Driving Style Classifier Training (Algorithm \ref{algo:pptrain})} \label{fig:algm2}
\end{figure}

 
Algorithm \ref{algo:pptrain} securely trains a decision tree out of the privately held distributed records. 
We conducted a group of experiments for classifier training (Algorithm \ref{algo:pptrain}) using $1000$, $2000$, $3000$, $4000$ and $5000$ drivers' traveling data featuring $9$ attributes. In each group of experiments, we tested the runtime for the encryption and decryption as well as the size of all the ciphertexts. 
As shown in Figure \ref{fig:algm2time} and \ref{fig:algm2size}, both the computational cost and the communication overheads present a linear increase trend as the number of vehicles increases.

 
 
 \subsection{Driving Style Recognition}
 
\begin{figure}[!htb]
	\centering \subfigure[Computational Cost]{
		\includegraphics[angle=0, width=0.9\linewidth]{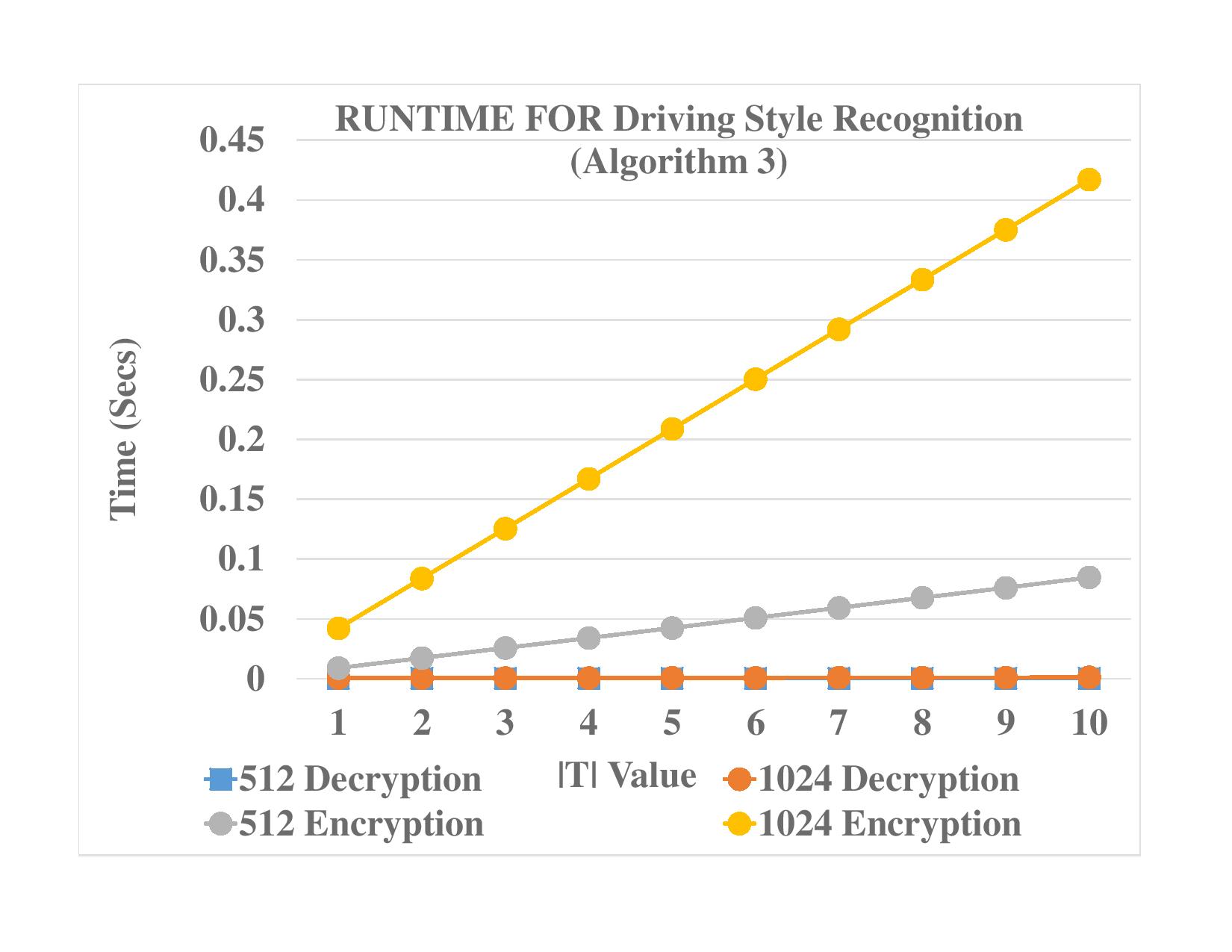}
		\label{fig:algm3time} } \subfigure[Communication Overheads]{
		\includegraphics[angle=0, width=0.8\linewidth]{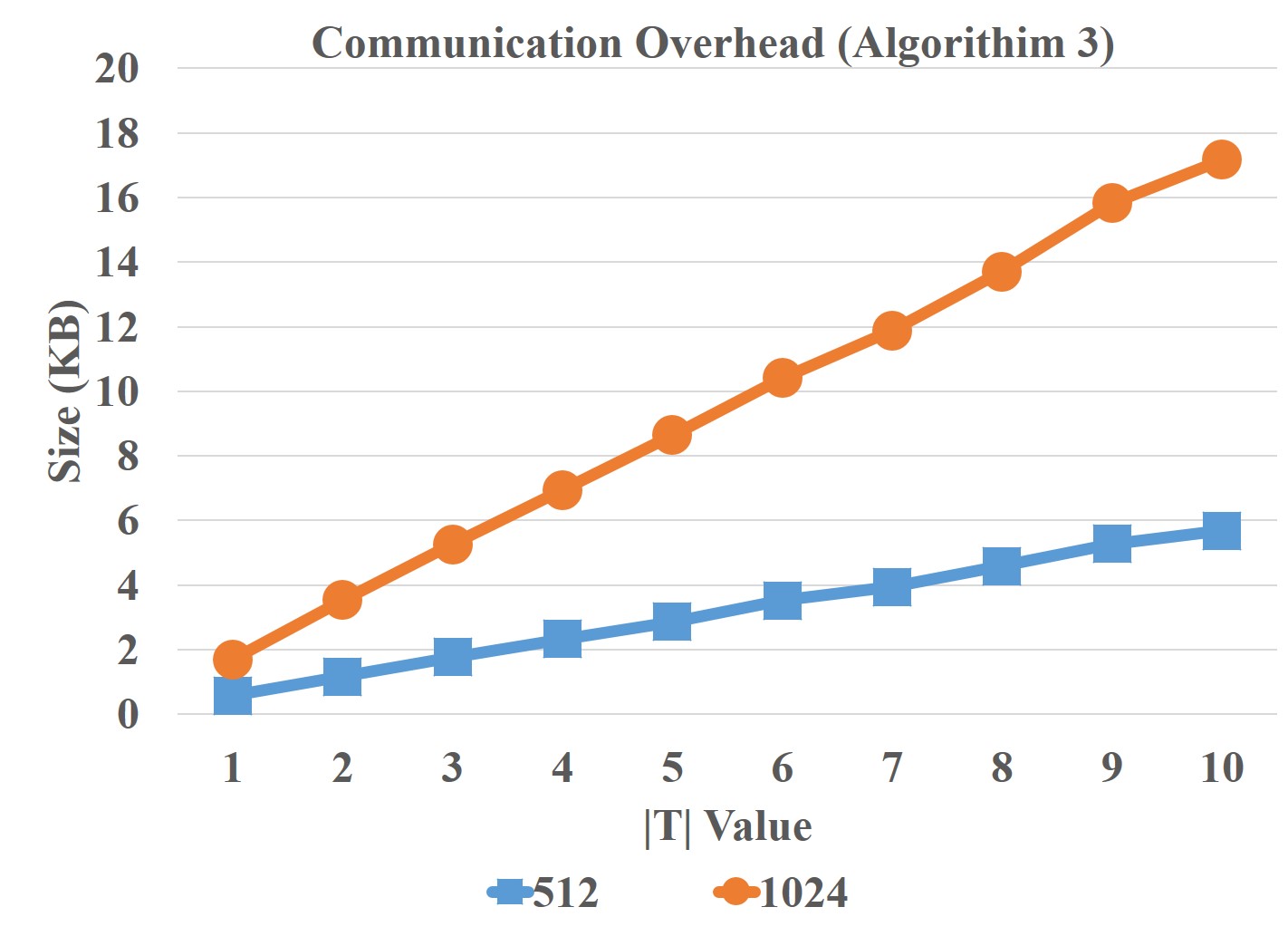}
		\label{fig:algm3size}  }
	\caption[Optional caption for list
	of figures]{Privacy Preserving Driving Style Recognition (Algorithm \ref{algo:ppreco})} \label{fig:algm3}
\end{figure}

Algorithm \ref{algo:ppreco} securely predicts the class for new individual drivers. Since the algorithm runs driving style recognition for all the drivers individually, we tested its computational costs and communication overheads again for a predicting a single driver's class. A group of experiments is conducted with a varying number of aggressive paths in the decision tree: $|T|=1,\dots,10$. As shown in Figure \ref{fig:algm3time} and \ref{fig:algm3size}, the costs increase slowly as the number of aggressive paths $|T|$ increases.

\section{Conclusion and Future Work}
\label{sec:concl}

In this paper, we have developed two secure communication protocols to tackle the privacy concerns in the two phases of driving style recognition among the insurance company and various vehicles. Participating parties can jointly train a decision tree based on the vehicles' historical traveling data without compromising their privacy. The insurance company can also use its decision tree to securely predict the driving style (aggressive or defensive) of any given driver with limited disclosure. We have also experimentally validated the performance of our proposed secure communication protocols.

In the future, we have several directions to extend this work. First, we assume a semi-honest adversarial model in this paper such that every participant will follow the outlined secure communication protocols.
In the real world, one or multiple parties may become more malicious to corrupt the protocol for additional payoff, or even collude with each other to breach privacy or jeopardize the utility of the protocol. We will explore efficient solutions to address the security and privacy concerns for multiparty driving style recognition in malicious adversarial model. Second, maybe more than one entities (e.g., multiple insurance companies and police department) would like to collaboratively predict the driving style of the drivers with their private inputs. Introducing more parties into this problem will influence the data partition scenario, and then the required secure communication protocols might be thoroughly different from the current ones. Third, in the future, vehicles' traveling data used for driving style recognition might be in real-time format rather than the historical aggregated format. In such scenario, the challenges on efficiency and bandwidth should be resolved. 

\end{document}